\documentclass[10pt, preprint]{emulateapj}

 \slugcomment{submitted to ApJ Supplement}

 \shorttitle{EUNIS-06 SW channel radiometric calibration and CDS/NIS calibration update}
 \shortauthors{Wang et al.}

\begin{document}

 \title{Absolute radiometric calibration of the EUNIS-06 170$-$205 \AA\ channel and calibration update for CDS/NIS}

 \author{Tongjiang Wang\altaffilmark{1,2}, Jeffrey W. Brosius\altaffilmark{1,2},
   Roger J. Thomas\altaffilmark{2}, Douglas M. Rabin\altaffilmark{2}, and Joseph M. Davila\altaffilmark{2}}

 \altaffiltext{1}{Department of Physics, Catholic University of America, 620 Michigan Avenue NE, 
    Washington, DC 20064, USA; wangtj@helio.gsfc.nasa.gov}
 \altaffiltext{2}{NASA Goddard Space Flight Center, Code 671, Greenbelt, MD 20771, USA}

\begin{abstract}
The Extreme-Ultraviolet Normal-Incidence Spectrograph sounding-rocket payload was flown on
2006 April 12 (EUNIS-06), carrying two independent imaging spectrographs covering
wave bands of 300$-$370~\AA\ in first order and 170$-$205~\AA\ in second order, 
respectively. The absolute radiometric response of the EUNIS-06 long-wavelength (LW) 
channel was directly measured in the same facility used to calibrate CDS prior to the $SOHO$ 
launch. Because the absolute calibration of the short-wavelength (SW) channel
could not be obtained from the same lab configuration, we here present a technique to derive it
using a combination of solar LW spectra and density- and temperature-insensitive line intensity ratios.
The first step in this procedure is to use the coordinated, cospatial EUNIS and $SOHO$/CDS
spectra to carry out an intensity calibration update for the CDS NIS-1 
waveband, which shows that its efficiency has decreased by a factor about 1.7 compared to that of the 
previously implemented calibration. Then, theoretical insensitive line ratios obtained from
CHIANTI allow us to determine absolute intensities of emission lines within the EUNIS SW bandpass from 
those of cospatial CDS/NIS-1 spectra after the EUNIS LW calibration correction. A total of 12 ratios 
derived from intensities of 5 CDS and 12 SW emission lines from Fe\,{\sc{x}}$-$Fe\,{\sc{xiii}} yield 
an instrumental response curve for the EUNIS-06 SW channel that matches well to a relative calibration 
which relied on combining measurements of individual optical components. Taking into account all
potential sources of error, we estimate that the EUNIS-06 SW absolute calibration is accurate 
to $\pm$20\%.
\end{abstract}


\section{Introduction}

The extreme-ultraviolet (EUV) waveband (150$-$1200 \AA) contains emission lines formed at temperatures ranging
from several times 10$^4$ K to several times 10$^7$ K and is therefore well suited for studies of the
multithermal structures in the solar atmosphere. The Solar EUV Research Telescope and Spectrograph (SERTS)
sounding rocket payload \citep{neu92} was developed at NASA's GSFC to study the Sun's corona and upper 
transition region with imaged EUV spectra. It was flown ten times, and produced excellent quality, 
high resolution spectra \citep[e.g.][]{tho94, bro96, bro97, bro98a, bro98b, bro99, bro00, and00}. 
Several SERTS flights have provided updated radiometric calibrations for the Coronal Diagnostic 
Spectrometer \citep[CDS;][]{har95} on the {\it Solar and Heliospheric Observatory} ($SOHO$) 
mission\citep[e.g.][]{tho02}.

 \begin{figure*}
 \epsscale{0.8}
 \plotone{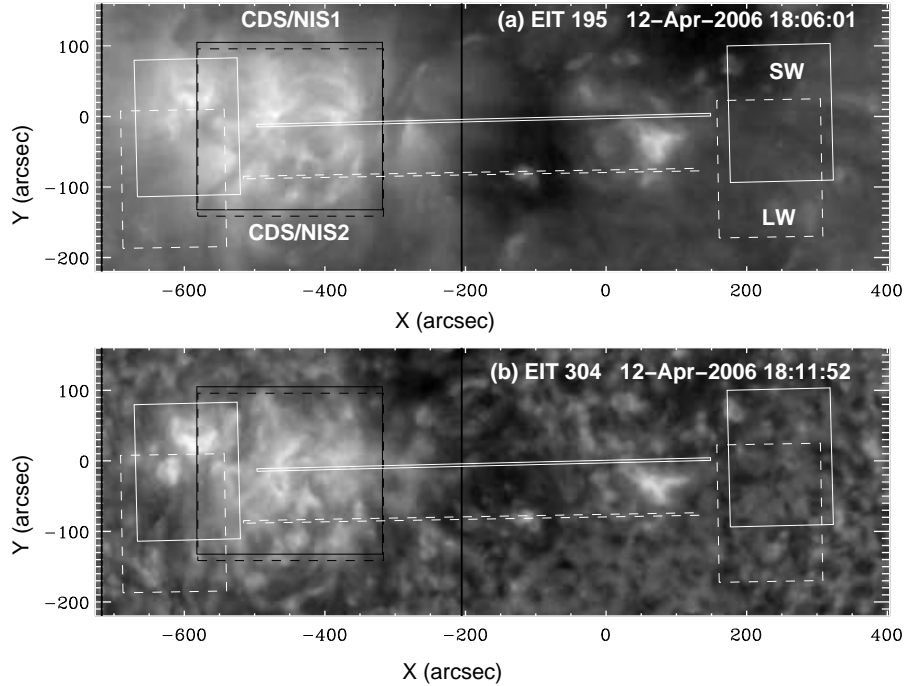}
 \caption{ \label{fgcln}
  Coalignments of the pointing for EUNIS-06 LW and SW channels with SOHO/CDS NIS-1 and NIS-2, SOHO/EIT, and TRACE.
The overlaid images are (a) EIT 195 \AA\ and (b) EIT 304 \AA.
The slit and two lobes positions for EUNIS SW and LW channels are marked with white solid and dashed lines,
respectively. The FOVs for CDS/NIS-1 and NIS-2 are marked with black solid and dashed lines, respectively.
Two vertical, black and thick solid lines mark the left and right boundaries of the FOV for TRACE.  }
 \end{figure*}
   
EUNIS (EUV Normal-Incidence Spectrograph) is a next-generation EUV imaging spectrograph that is 100 times 
more sensitive than SERTS, with $\sim$5${''}$ spatial resolution and $\sim$2~s time cadence \citep{tho01}. 
This powerful instrument can pave the way for entirely new studies of the corona using high cadence spectra 
at a single location, rapid rastering over large 2D areas of the solar surface, or deep exposures on 
intrinsically faint coronal features.  EUNIS has now been successfully flown twice, in April 2006 and 
November 2007 \citep{rab08}. Scientific results have already been reported using the
EUNIS-06 data. For example, \citet{bro07} found both upflows and downflows in a coronal bright point observed 
in emission lines formed at $T$=0.05$-$2.5 MK, providing evidence for magnetic reconnection in bright points.
\citet{bro08a} further derived coronal electron densities, the differential emission measure, and elemental 
abundances of this bright point from line intensities. \citet{bro08b} studied a cool transient event seen only
in He\,{\sc{ii}} with high-cadence spectra, revealing relative upflows up to 20 km~s$^{-1}$
which are consistent with gentle chromospheric evaporation. \citet{jes08} found short-period (26$\pm$4 s) velocity 
oscillations in the He\,{\sc{ii}} 303.8 \AA\ line and other transition region (TR) lines formed at temperatures
up to 4$\times$10$^5$ K and suggested an interpretation in terms of fast sausage modes in a flux tube.

For EUNIS-06, a complete end-to-end radiometric calibration of its long wavelength (LW) channel was carried out 
at the Rutherford Appleton Laboratory (RAL) in England using a stable EUV transfer light source provided by
the German Physikalisch-Technische Bundesanstalt (PTB). Unfortunately, it was not possible to obtain a similar
measurement of the EUNIS short-wavelength (SW) channel due to the weakness of available lines in that
waveband. Therefore, we have developed a technique for deriving the absolute radiometric calibration of 
its SW channel by combining measured LW solar spectra with density- and temperature-insensitive line intensity 
ratios. The procedure of using insensitive line intensity ratios to derive relative radiometric calibration was
first suggested by \citet{neu83} and has been successfully applied in calibrations of SERTS and other instruments.

 \begin{figure}
 \epsscale{1.0}
 \plotone{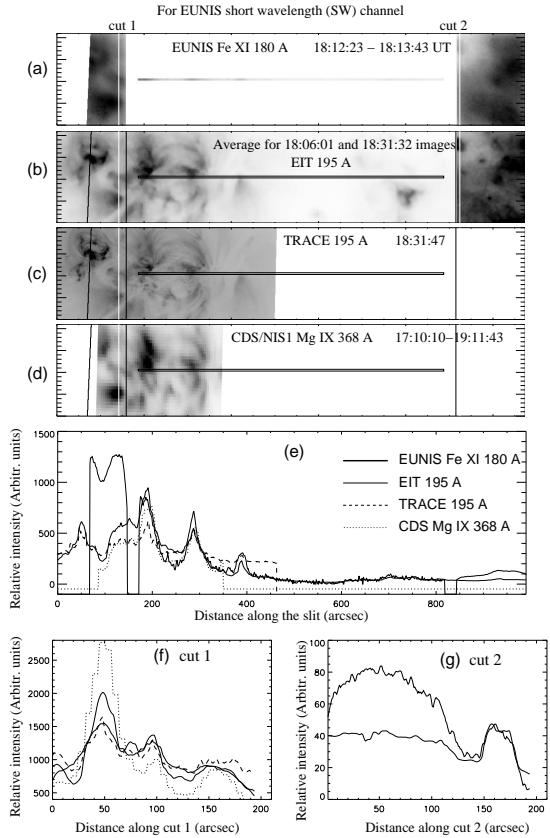} 
 \caption{\label{fgswmg}
Comparisons of images and intensity profiles from observations of different instruments. (a) Image of Fe\,{\sc{xi}} 
180 \AA\ line intensity from the EUNIS-06 SW channel. The EUNIS SW slit is the nearly-horizontal line toward 
the center of panels (a)-(d). (b) The EIT 195 \AA\ image, in which the 
contrast in the right lobe is enhanced in order to be seen clearly. (c) The TRACE 195 \AA\ image. 
(d) The Mg\,{\sc{ix}} 368 \AA\ line intensity image from CDS/NIS-1.(e) Comparisons of intensity profiles
along the EUNIS slit (including its extending in two lobes). (f) and(g) Comparisons of intensity profiles
along cut1 and cut2 (white vertical lines) marked in (a), respectively. }
 \end{figure}

\section{Observations and Data reduction}
EUNIS-06 was launched from White Sands Missile Range, New Mexico, at 18:10 UT on 2006 April 12. It achieved
a maximum altitude of 313 km and obtained solar spectra and images between 1812 and 1818 UT. EUNIS observed
NOAA AR 10871 (S07$^{\circ}$, E28$^{\circ}$) and AR 10870 (S08$^{\circ}$, W02$^{\circ}$), the quiet region
between them and a southern-hemisphere coronal bright point within the quiet area \citep{bro07, bro08b}. Coordinated
observations were obtained with the EUV Imaging Telescope \citep[EIT;][]{del95} and the CDS aboard $SOHO$, 
as well as with the {\it Transition Region and Coronal Explorer} \citep[$TRACE$;][]{han99}. 
We coaligned the data from different instruments with a cross-correlation method, taking the EIT images 
as a reference.

\subsection{EUNIS}
The optical design of EUNIS is directly based on that of SERTS, but features two independent, imaging 
spectrographs, one covering EUV lines between 300 and 370 \AA\ seen in first order (LW channel), and 
a second covering lines between 170 and 205 \AA\ seen in second order (SW channel). For both channels,
the optical design of telescope, entrance slit, toroidal grating, and detector are identical, with the
only difference being the multilayer coatings applied to the reflecting surfaces. Each channel includes
a microchannel-plate (MCP) intensifier section and a set of three 1K$\times$1K active pixel sensors (APS), 
which are coupled to the MCP by fine-pore fiber-optic blocks. Areas of the solar image selected by the two
entrance apertures include a narrow 2$^{''}\times$660$^{''}$ slit along which spatially resolved spectra are
obtained, and wider ``lobes'' at the ends from which spectroheliograms are obtained. For this flight, 
the lobe-slit fields of view were oriented East-West near the center of the solar disk. EUNIS has a first-order 
spectral dispersion of 25 m\AA\ pixel$^{-1}$ and a designed spatial scale of 0$^{''}$.927 pixel$^{-1}$. 
A more detailed description of EUNIS design was given by \citet{tho01}.
For the first flight of EUNIS, its optics limited the actual spatial resolution to about 5$^{''}$ 
and the measured spectral resolution was $\sim$200 and $\sim$100 m\AA\ FWHM in the LW and SW channels.

In this first flight we obtained 80 solar exposures in ``stare'' (fixed pointing) mode at the start of 
the flight with $\sim$2 s cadence for all but the first five; further observations were made in scanning 
(spectroheliogram) mode. All of the raw data were processed with several routine adjustments, including 
dark image subtraction, flat-fielding and non-linearity correction. The quality of our EUNIS data is 
sufficient to reveal non-linearities in the APS response that reach about 15\% as they approach saturation. 
This effect is identical for all APS units, and has been characterized by ratios of different exposure times 
to a constant EUV source in the laboratory. Appropriate corrections for the effect are applied to raw data to 
convert the recorded Data Numbers (DN) into Relative Exposure Units (REU), which are then used in all subsequent
analyses. We used 34 exposures of the stare sequence to create an averaged lobe image for 
coalignments and an averaged spectral image for calibration.

 \begin{figure}
 \epsscale{1.0}
 \plotone{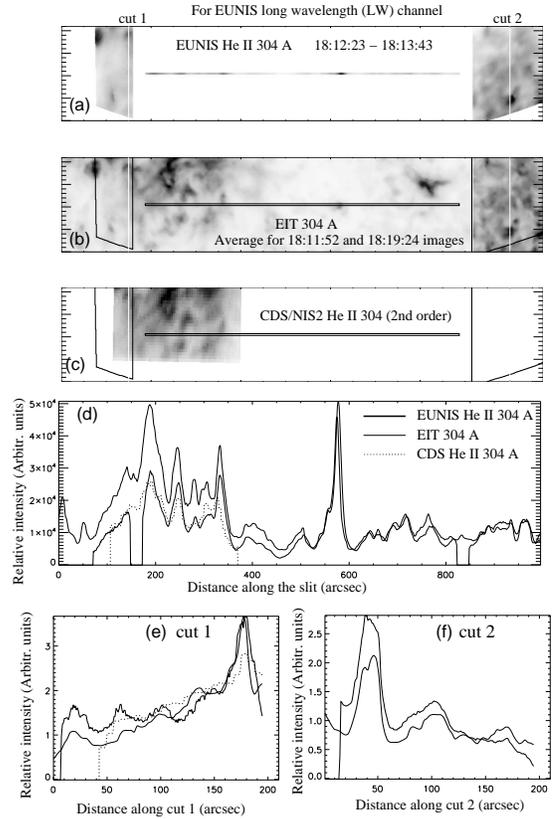}
 \caption{\label{fglwmg}
Comparisons of images and intensity profiles from observations of different instruments. (a) He\,{\sc{ii}} 
304 \AA\ line intensity image from the EUNIS-06 LW channel. The nearly-horizontal line in panels (a)-(c) 
represents the EUNIS LW slit. (b) The EIT 304 \AA\ image, in which the 
contrast in the right lobe is enhanced in order to be seen clearly. (c) Image of the He\,{\sc{ii}} 304 \AA\ 
(second order) line intensity from CDS/NIS-2. (d) Comparisons of intensity profiles
along the EUNIS slit (including its extendings in two lobes). (e) and(f) Comparisons of intensity profiles
along cut1 and cut2 (white vertical lines) marked in (a), respectively. }
 \end{figure}

\subsection{SOHO/CDS}
The CDS includes a Normal Incidence Spectrometer (NIS) that can be used to obtain
stigmatic EUV spectra within its 308$-$381 \AA\ (NIS-1) and 513$-$633 \AA\ (NIS-2)
wavebands along its 240$^{''}$ long slit. Several slit widths are available, the most commonly
used being 4$^{''}$. The instrument can be operated in a sit-and-stare mode wherein successive
exposures over small (single slit) areas are obtained, or it can be used to obtain raster
images of target areas by obtaining spectra from successive slit pointings. In the latter case, 
the CDS scans a region of the Sun from the West to East. Since the CDS pointing has no compensation
for solar rotation during observations, the actual field-of-view (FOV) of an obtained spectroheliogram
will be stretched out in the x-direction when the targeted region is located on the solar disk. 
In this study, the CDS rastering observation was made during 17:10:10$-$19:11:43 UT, consisting of 
60 pointing positions. The FOV is 264$^{''}\times237^{''}$. The pointing difference between 
CDS NIS-1 and NIS-2 has been corrected when applying {\it mk\_cds\_map} in the Solar Software (SSW)
IDL library.

 \begin{figure*}
  \epsscale{1.0}
 \plotone{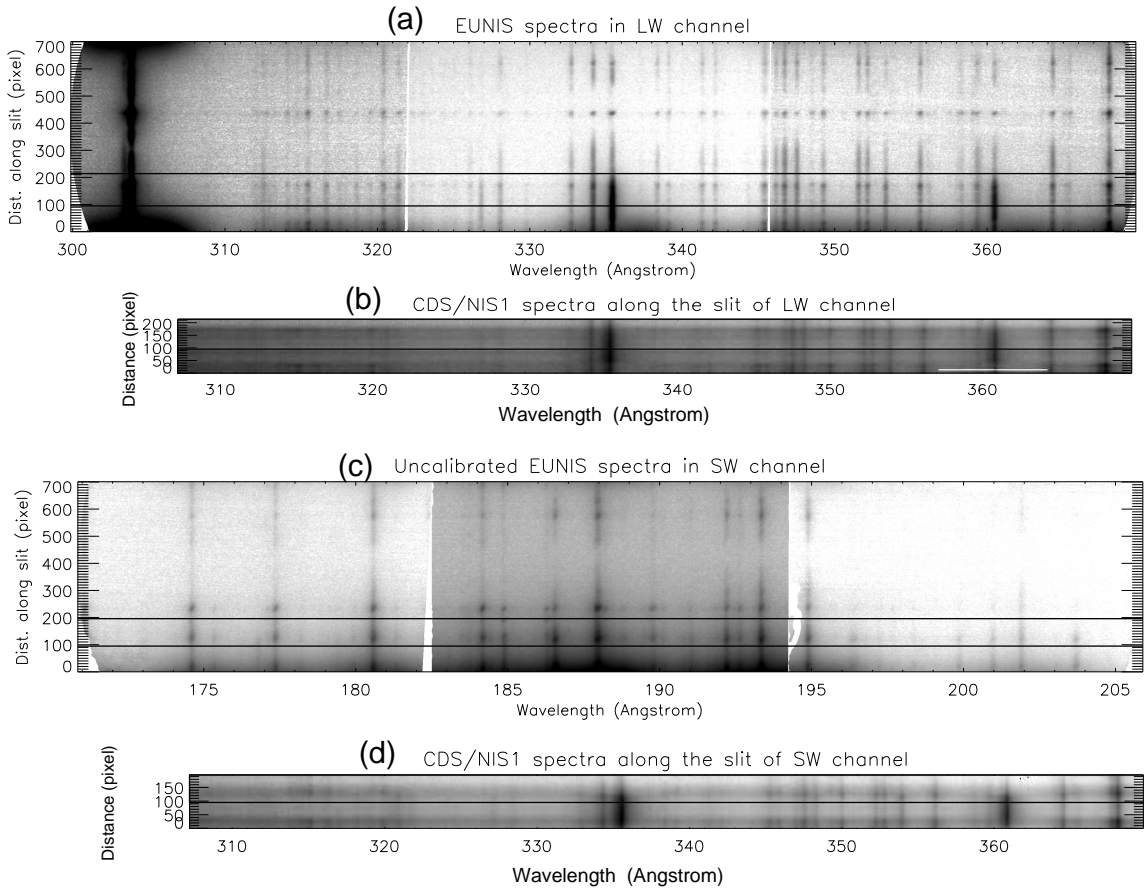}
 \caption{\label{fgspmg}
 (a) EUNIS-06 full window spectra along the slit of LW channel, averaged for the data observed during 18:12:23 $-$
18:13:43 UT. (b) CDS/NIS-1 spectra extracted along the slit position of EUNIS LW. 
(c) EUNIS full window spectra along the slit of SW channel, averaged for the data observed during 18:12:23 $-$
18:13:43 UT. (d) CDS/NIS-1 full window spectra extracted along the slit position of EUNIS SW. In all panels, 
the spectra are shown as negative images, where darker areas correspond to greater intensities. 
The y-axis in all plots is in units of EUNIS pixels. In (a) and (b), two horizontal lines mark the same section
(111${''}$ wide) along the LW slit. In (c) and (d), two horizontal lines mark the same section (94${''}$ wide)
along the SW slit.  The EUNIS and CDS spectra averaged in these sections are used to measure intensities
of the lines used for radiometric calibration.}
 \end{figure*}

\subsection{Coalignments}
We used the EUNIS lobe images to precisely coalign EUNIS data with full disk solar images obtained
with EIT. We first scaled the EIT images to the same pixel size as EUNIS. We then
coaligned the left lobe of an EUNIS image with the EIT image by applying a cross-correlation technique. 
If the EUNIS image (or slit) is tilted to the East-West direction and its actual spatial
scale is different from the design value, its right lobe will have offsets to the EIT image that 
has been coaligned by the left lobe. So we can estimate the tilt angle of the EUNIS slit and actual spatial
resolution from the measurements of these offsets. For the EUNIS SW channel, we measured the offsets of
$dx$=4.7 pixel in the $x$-direction and $dy$=$-$20.6 in the $y$-direction for the right lobe by comparing 
EUNIS Fe\,{\sc{xi}} $\lambda$180 image with the EIT $\lambda$195 image. We then determined the actual 
pixel size from flight data by 
$R_{\rm flight}\approx R_{\rm design}D/(D+dx)$     
where $R_{\rm flight}$ and $R_{\rm design}$ are the actual (from flight measurements) and designed (nominal) 
pixel sizes of EUNIS images, and $D$ is the distance in EUNIS pixels between two lobe regions used for 
coalignments. The tilt angle of the EUNIS image (or slit) can be determined by
$\theta\approx{\rm tan}^{-1}[dy/(D+dx)]$.
From the solar measurements we obtained $R_{\rm flight}$=0$^{''}$.922~pixel$^{-1}$ and $\theta$=1.4$^{\circ}$
counterclockwise from the East-West. Similarly for the LW channel, we measured the offsets of the right
lobe by comparing EUNIS He\,{\sc{ii}} $\lambda$304 image with the EIT $\lambda$304 image, and obtained
$R_{\rm flight}$=0$^{''}$.927~pixel$^{-1}$ and $\theta$=1.0$^{\circ}$.
The measurements also show that the EUNIS-06 SW and LW slits were not co-pointing in flight, but were
instead separated by about 76$^{''}$ in the $y$-direction (Fig.~\ref{fgcln}). 

We coaligned the CDS images by comparing NIS-1 Mg\,{\sc{ix}} $\lambda$368, Si\,{\sc{x}} $\lambda$347,
Fe\,{\sc{xii}}/Fe\,{\sc{xi}} $\lambda$352 images with the EIT $\lambda$195 image, and comparing NIS-2 
He\,{\sc{i}} $\lambda$584 image with the EIT $\lambda$304 image. The coalignment between CDS and EUNIS
was made based on the 
average offsets measured above. Figure~\ref{fgcln} shows positions of the FOVs for the coaligned EUNIS
LW and SW channels, CDS NIS-1 and NIS-2, and TRACE relative to the EIT images. Figures~\ref{fgswmg}
and \ref{fglwmg} show the EUNIS SW  and LW channel images and their comparisons with the coaligned EIT,
TRACE and CDS/NIS images. A good agreement in their intensity profiles along two cuts in the lobe regions 
and along the EUNIS slit indicate that the images from the different instruments have been well coaligned.

 \begin{figure*}
  \epsscale{1.0}
 \plotone{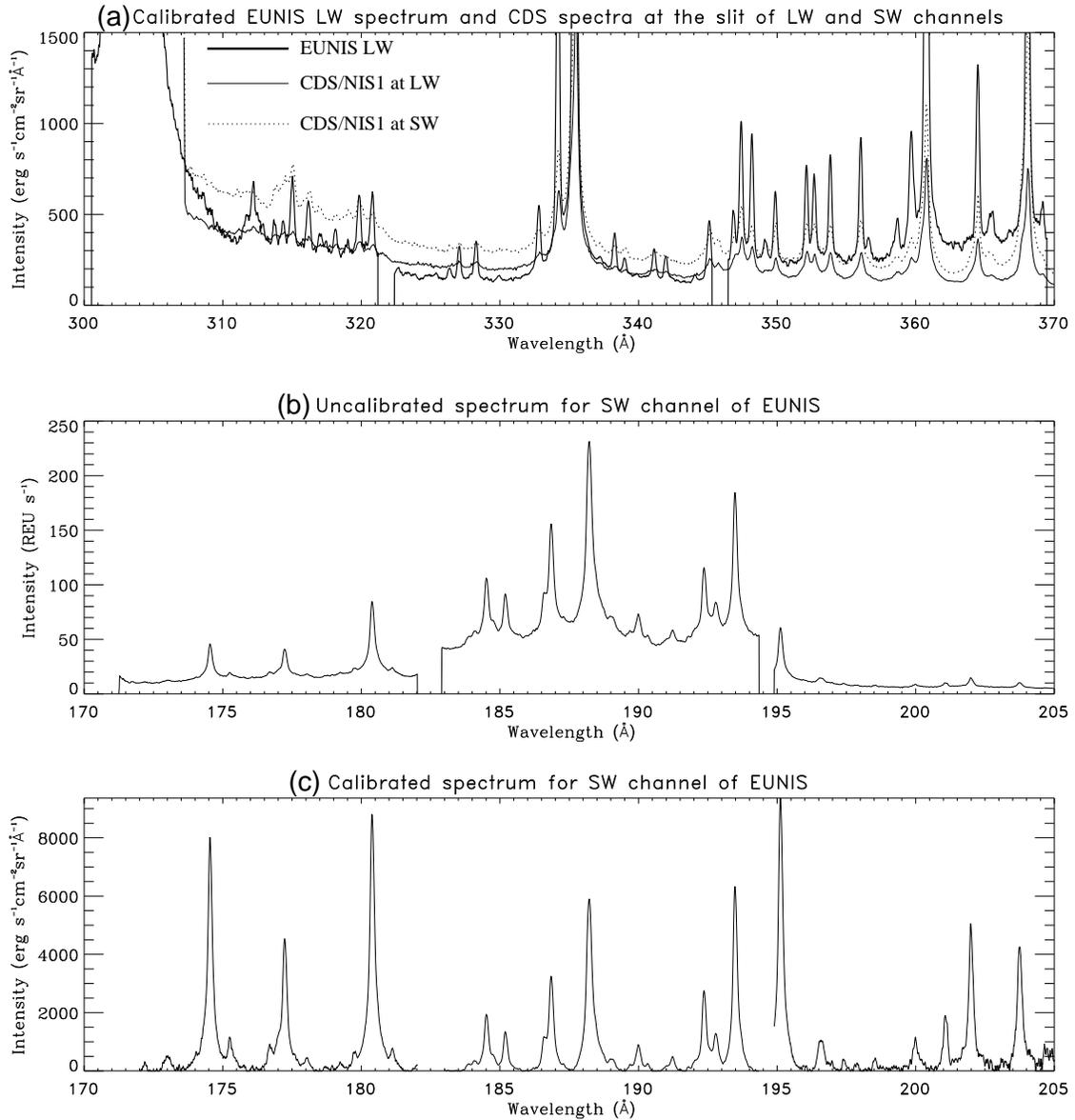}
 \caption{\label{fgspfl}
 (a) Calibrated EUNIS/LW channel (thick solid line) and CDS/NIS-1 (thin solid line) spectra, spatially 
averaged over a segment along the LW slit shown in Fig.~\ref{fgspmg}a and b. The CDS/NIS-1 spectrum, 
averaged over a segment along the SW slit shown in Fig.~\ref{fgspmg}d, is also plotted 
(dotted line) for comparison. (b) The uncalibrated EUNIS/SW channel spectrum, spatially averaged over 
a segment shown in Fig.~\ref{fgspmg}c. (c) Same as that shown in (b) but after background subtraction and
application of the absolute radiometric calibration shown in Fig.~\ref{fgswcl}b. }
 \end{figure*}

 \begin{deluxetable*}{rlcccc}
 \tablecaption{  EUNIS-06 LW channel calibration verification from density- and temperature-insensitive line ratios. 
\label{tablwr}}
 \tablewidth{0pt}
 \tablehead{
 \colhead{Ion} &  \colhead{Wavelength} & \colhead{Theo. Rel.} & \colhead{Observed Intensity} & \colhead{Observed Rel.} & Normalized\\
  & \colhead{(\AA)} & \colhead{Intensity} & \colhead{(ergs cm$^{-2}$s$^{-1}$sr$^{-1}$)} & \colhead{Intensity} & \colhead{Col.(5)/Col.(3)}\\
 \colhead{(1)} & \colhead{(2)} & \colhead{(3)} & \colhead{(4)} & \colhead{(5)} & \colhead{(6)}}   
 \startdata

Mg\,{\sc{viii}}.....
& 313.75   & 0.357$\pm$ 0.021   &  28.53$\pm$   2.85   & 0.252$\pm$ 0.036   & 0.823$\pm$ 0.126\\
& 315.04   & 1.000$\pm$ 0.000   & 113.09$\pm$  11.31   & 1.000$\pm$ 0.100   & 1.165$\pm$ 0.116\\
& 317.04   & 0.252$\pm$ 0.015   &  25.17$\pm$   2.67   & 0.223$\pm$ 0.032   & 1.029$\pm$ 0.162\\
& 339.01   & 0.229$\pm$ 0.005   &  18.40$\pm$   6.03   & 0.163$\pm$ 0.056   & 0.828$\pm$ 0.284\\
\\
Si\,{\sc{viii}}.....
& 314.33   & 0.336$\pm$ 0.000   &  30.84$\pm$   3.08   & 0.333$\pm$ 0.047   & 0.946$\pm$ 0.134\\
& 316.21   & 0.670$\pm$ 0.000   &  80.97$\pm$   8.10   & 0.873$\pm$ 0.123   & 1.245$\pm$ 0.176\\
& 319.83   & 1.000$\pm$ 0.000   &  92.73$\pm$   9.27   & 1.000$\pm$ 0.100   & 0.955$\pm$ 0.096\\
\\
Si\,{\sc{ix}}.......
& 341.99   & 0.362$\pm$ 0.030   &  33.66$\pm$   6.37   & 0.347$\pm$ 0.074   & 0.964$\pm$ 0.221\\
& 345.12\tablenotemark{a}   & 1.000$\pm$ 0.000   &  97.13$\pm$   9.71   & 1.000$\pm$ 0.100   & 1.007$\pm$ 0.101\\
\\
Fe\,{\sc{xi}}.......
& 341.11   & 0.281$\pm$ 0.039   &  44.57$\pm$   6.55   & 0.334$\pm$ 0.059   & 1.161$\pm$ 0.262\\
& 352.66   & 1.000$\pm$ 0.000   & 133.61$\pm$  13.36   & 1.000$\pm$ 0.100   & 0.978$\pm$ 0.098\\
\\
Fe\,{\sc{xii}}......
& 346.85   & 0.304$\pm$ 0.002   &  75.19$\pm$   8.68   & 0.276$\pm$ 0.042   & 0.969$\pm$ 0.148\\
& 352.11   & 0.604$\pm$ 0.011   & 142.45$\pm$  14.25   & 0.523$\pm$ 0.074   & 0.924$\pm$ 0.132\\
& 364.47   & 1.000$\pm$ 0.000   & 272.14$\pm$  27.21   & 1.000$\pm$ 0.100   & 1.066$\pm$ 0.107\\
\\
Fe\,{\sc{xvi}}......
& 360.761   & 0.480$\pm$ 0.000   &1299.37$\pm$ 129.94   & 0.512$\pm$ 0.072   & 1.046$\pm$ 0.148\\
& 335.410   & 1.000$\pm$ 0.000   &2536.80$\pm$ 253.68   & 1.000$\pm$ 0.100   & 0.980$\pm$ 0.098\\
 \enddata
\tablenotetext{a}{Si\,{\sc{ix}} 345.12 was self-blended with 344.95.}
\end{deluxetable*}

\section{Radiometric calibration of EUNIS LW channel}
\label{sctlwc}
Direct calibration of the EUNIS-06 LW channel was carried out in August 2006 
at RAL in the same facility and using the same EUV light source as was used for pre-flight calibrations 
of CDS \citep{lan02} and the EUV Imaging Spectrometer (EIS) \citep{cul07} on $Hinode$. Recalibration of 
the PTB light source against the primary EUV radiation standard of BESSY-II in March 2007 showed that it
had remained stable within its 7\% uncertainty. Measurements were made of the He\,{\sc{ii}} 304 \AA\ line 
and of 11 distinct Ne features between 300 and 370 \AA\ at 157 individual locations covering the instrument's 
entrance aperture. The aperture-averaged response at each wavelength was combined with the known source 
flux and with geometric factors, such as aperture area and pixel size, resulting in the absolute EUNIS 
response within a total uncertainty of 10\% over its full LW bandpass \citep{tho08}.

\begin{figure}
  \epsscale{1.0}
 \plotone{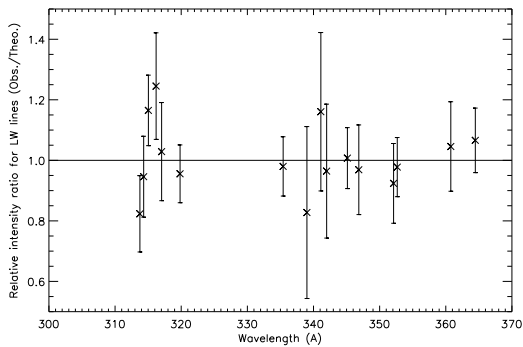}
 \caption{\label{fgtblw}
 Plot of the observed-to-theoretical intensity ratios (col.[6] of Table~\ref{tablwr}), normalized to the
weighted average ratio within each line group, as a function of wavelength for the EUNIS LW channel.}
\end{figure}

We validated the laboratory calibration of the EUNIS LW channel by checking its relative 
calibration using the density- and temperature-insensitive line ratio method \citep{bro98a, bro98b}. 
Table~\ref{tablwr} lists six groups of emission lines for Mg\,{\sc{viii}}, Si\,{\sc{viii}}, 
Si\,{\sc{ix}}, Fe\,{\sc{xi}}, Fe\,{\sc{xii}}, and Fe\,{\sc{xvi}}. 
Column (3) gives the theoretical line intensity relative
to the strongest one in each group, which were calculated with the CHIANTI (version 6) package 
\citep{der97, der09}. The line ratios exhibit slight variations with electron density ($N_e$), so we
took the mean value over a range of $8.5<{\rm log}_{10}N_e<10.5$, and the uncertainty corresponds
to half of the difference between the maximum and minimum values. In each case, the intensity ratios were calculated 
at the temperature of maximum ion abundance (i.e., the lines' formation temperature), using the ionization 
equilibrium data ({\it chianti.ioneq}). Column (4) gives the observed intensities of calibrated spectral lines
measured from the averaged spectra (see Fig.~\ref{fgspmg}a and Fig.~\ref{fgspfl}a), 
and column (5) the corresponding relative intensities to the same lines used in column (3) 
within each line group. Ideally, the values listed in column (5) should
match those listed in column (3). Column (6) gives the ratio of the observed relative intensities to the theoretical ones,
normalized by the weighted average ratio within each group. Ratios close to unity indicate good agreement between
the observed and the theoretical values. Column (6) in Table~\ref{tablwr} reveals that all but
three of the ratios agree to within one standard deviation (1$\sigma$) of measurements and that all of 
them agree within factors
better than 2. This is illustrated in Figure~\ref{fgtblw}, where the values in column (6) are plotted as a
function of wavelength. The fact that many line intensity ratios from different elements and ionization stages
yield mutually consistent results supports both the general validity of the atomic physics
calculations and our laboratory calibration of the EUNIS LW channel at RAL.

 \begin{figure}
  \epsscale{1.0}
 \plotone{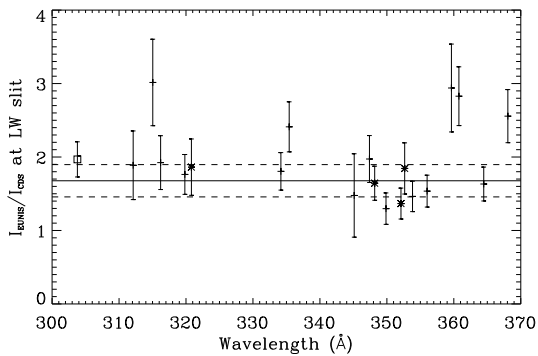}
 \caption{\label{fgeucd}
  Plot of the EUNIS/LW-to-CDS/NIS-1 intensity ratios for the spectral lines listed in Table~\ref{tablin}. 
The solid line represents the average ratio for those lines with the ratios less than 2 and the dashed lines
their standard deviation. Intensity ratios in excess of 2 are discussed in the text. The data points denoted 
with  {\it asterisks} indicate members of temperature- and density-insensitive line groups used for 
calibration of the EUNIS SW channel. The point denoted with a {\it box} corresponds to the He\,{\sc{ii}} 
303.78 \AA\ line of CDS NIS-2 in second order. }
 \end{figure}

\section{CDS underflight calibration with EUNIS LW}
The lab-calibrated SERTS-97 was successfully used to improve the sensitivity curve for the 
CDS NIS-1 waveband based on their coordinated, cospatial spectra \citep{tho02}. In the
following, we present a new calibration update for CDS NIS-1 based on the EUNIS LW observation, and then
apply this updated CDS response to calibrate the EUNIS SW channel in the next section.

Figures~\ref{fgspmg}a and \ref{fgspmg}b show the time-averaged EUNIS LW spectra and the cospatial CDS NIS-1 spectra. 
CDS NIS-1 and EUNIS made coordinated observations of AR 10871, which was located near the eastern 
end of the EUNIS slit. To reduce the effect of scattered light from the EUNIS lobe, we averaged the spectra 
along the slit over an area away from that lobe. Figure~\ref{fgspfl}a shows a comparison between the 
averaged spectra in the wavelength range from 300 to 380 \AA, which clearly
indicates that EUNIS has much higher sensitivity and spectral resolution than CDS. Especially
for spectral lines of CDS in the wavelength range 310$-$330 \AA, the measurement of line intensities 
is challenging because the lines are weak and not well separated. We fitted the EUNIS spectral lines with 
Gaussian functions and fitted the CDS lines with broadened Gaussian functions \citep{tho99} using 
the standard SSW line profile fitting routine ({\it xcfit.pro}). \citet{tho99} found that the broadened 
Gaussian function appears to well match the line profiles of the post-recovery CDS data 
after the attitude-loss of SOHO in 1998. 

Table~\ref{tablin} lists 20 emission lines observed in both EUNIS LW and CDS (NIS-1 in first order and NIS-2 
in second order) and their measured intensities with a 1$\sigma$ uncertainty. For CDS, absolute
intensities are based on the previously most recent rocket underflight calibrations implemented on 28 Feb 2000
based on EGS \citep[the EUV Grating Spectrometer;][]{woo98} and SERTS-97 \citep{lan02, tho02}. (Out of these lines, 
four were selected to calibrate the EUNIS SW bandpass based on their membership in insensitive line groups
that include lines scattered throughout the SW bandpass, as described below.) The He\,{\sc{ii}} $\lambda$304 
line is a second order line of NIS-2. Since in the EUNIS LW bandpass the scattered light from the 
lobe image around this line is very strong (Fig.~\ref{fgspmg}a), we chose a smaller area from pixel No. 142 
to 214 along the slit to reduce this effect as much as possible. Note that Si\,{\sc{xi}} $\lambda$303 is 
another NIS-2 second-order line, but as it is too weak to be reliably separated from the strong He\,{\sc{ii}} 
line in the CDS spectrum, we did not include its measurements here. The EUNIS LW to CDS line intensity ratios 
are given in column 5 of Table~\ref{tablin} and shown in Figure~\ref{fgeucd}. We suspect that the CDS sensitivity 
in the spectral vicinities of the strongest lines in our spectra (Fe\,{\sc{xvi}} at 335.4 and 360.8 \AA,
Mg\,{\sc{ix}} at 368.1, and Mg\,{\sc{viii}} at 315.0 \AA) are reduced even more than the typical value $\lesssim$2 
due to detector burn-in by persistent exposure to these strong lines. For this reason we averaged over only those 
lines with the EUNIS-to-CDS line ratios less than 2 to obtain their relative calibration factor, which is 1.68 $\pm$0.22.
This factor is a correction to the previous CDS calibration, which indicates that the response efficiency of CDS NIS-1 
has decreased by a factor of about 1.7 in the nine years between absolute calibration updates. 
The measurements also indicate that the response efficiency of CDS NIS-2 in the second order at 303$-$304 \AA\ has 
decreased by a factor of about 2 during this time interval. These results are in very good agreement with a recent 
radiometric calibration correction for CDS NIS by \citet{zan10} who also found a factor of about 2 decrease in
responsivity for most wavelengths over 10 years.

 \begin{deluxetable*}{llccc}
 \tablecaption{ Absolutely calibrated active region line lists for EUNIS LW and CDS NIS-1/NIS-2  bandpasses
 \label{tablin}}
 \tablewidth{0pt}
 \tablehead{
  \colhead{Wavelength}\tablenotemark{a} & \colhead{Ion} & \colhead{EUNIS Intensity} & \colhead{CDS Intensity} & \colhead{EUNIS-to-CDS}\\
  \colhead{(\AA)} &   & \colhead{(ergs cm$^{-2}$s$^{-1}$sr$^{-1}$)} & \colhead{(ergs cm$^{-2}$s$^{-1}$sr$^{-1}$)} & \colhead{Intensity Ratio}}
  \startdata
   303.78$^{\diamond}$...... &  He\,{\sc{ii}} & 14272.1$\pm$1427.2  & 7253.1$\pm$ 502.3   &1.97$\pm$0.24\\
   312.11 ....... & Fe\,{\sc{xiii}} & 90.5 $\pm$9.1 & 48.0$\pm$10.9  & 1.89$\pm$0.47\\
   315.04 ....... & Mg\,{\sc{viii}} & 113.1$\pm$11.3 & 37.5$\pm$6.3  &3.01$\pm$0.59\\
   316.21 ....... & Si\,{\sc{viii}} &  81.0$\pm$8.1 &  42.1$\pm$6.8  &1.92$\pm$0.37\\
   319.83 ....... & Si\,{\sc{viii}} &  92.7$\pm$9.3  & 52.6$\pm$6.2  &1.76$\pm$0.27\\
   320.81$^*$...... & Fe\,{\sc{xiii}}& 98.1$\pm$9.8 & 52.7$\pm$9.5   &1.86$\pm$0.38\\ 
   334.17 ....... & Fe\,{\sc{xiv}}  & 586.7$\pm$58.7 & 325.1$\pm$32.5 &1.80$\pm$0.26\\
   335.41 ....... & Fe\,{\sc{xvi}}\tablenotemark{b}  &2714.2$\pm$271.4 & 1125.9$\pm$112.6 &2.41$\pm$0.34\\
   345.12 ....... & Si\,{\sc{ix}}\tablenotemark{c}   & 97.1$\pm$9.7 & 65.8$\pm$24.4       &1.48$\pm$0.57\\
   347.40 ....... & Si\,{\sc{x}}    & 222.7$\pm$22.3  & 112.9$\pm$14.4  &1.97$\pm$0.32\\
   348.18$^*$...... & Fe\,{\sc{xiii}} & 194.1$\pm$19.4  & 118.1$\pm$11.8 &1.64$\pm$0.23\\
   349.87 ....... & Si\,{\sc{ix}}   & 103.2$\pm$10.3  & 79.6$\pm$10.4 &1.30$\pm$0.21\\
   352.11$^*$...... & Fe\,{\sc{xii}} & 142.5$\pm$14.3 & 104.3$\pm$12.2 &1.37$\pm$0.21\\
   352.66$^*$...... & Fe\,{\sc{xi}}& 133.6$\pm$13.4  & 72.4$\pm$11.6 &1.84$\pm$0.35\\
   353.83 ....... & Fe\,{\sc{xiv}}  & 150.8$\pm$15.1  & 103.0$\pm$10.3 &1.46$\pm$0.21\\
   356.01 ....... & Si\,{\sc{x}}    & 184.5$\pm$18.5 & 120.2$\pm$12.0  &1.54$\pm$0.2\\
   359.64...... & Fe\,{\sc{xiii}}\tablenotemark{d} & 168.3$\pm$16.8 & 57.3$\pm$10.1 &2.94$\pm$0.60\\
   360.76 ....... & Fe\,{\sc{xvi}}  & 1299.4$\pm$130.0 & 459.5$\pm$46.0 &2.83$\pm$0.40\\
   364.47...... & Fe\,{\sc{xii}}& 272.1$\pm$27.2 & 166.7$\pm$16.7  &1.63$\pm$0.23\\
   368.07 ....... & Mg\,{\sc{ix}}\tablenotemark{e}   & 902.8$\pm$90.3 & 353.1$\pm$35.3 &2.56$\pm$0.36\\
  \enddata
 \tablenotetext{a}{The line wavelength data are taken from CHIANTI line list. Those lines marked with 
{\it asterisks} are members of temperature- and density-insensitive line groups which were used to calibrate the EUNIS
SW bandpass. The line marked with a {\it diamond} is a second order line in CDS NIS-2.}
 \tablenotetext{b}{Fe\,{\sc{xvi}} 335.41 was blended with Mg\,{\sc{viii}} 335.25.}
 \tablenotetext{c}{Si\,{\sc{ix}} 345.12 was self-blended with 344.95.}
 \tablenotetext{d}{Fe\,{\sc{xiii}} 359.64 was self-blended with 359.84.}
 \tablenotetext{e}{Mg\,{\sc{ix}} 368.07 was blended with Mg\,{\sc{vii}} 367.67 and 367.68.}
\end{deluxetable*}

\begin{figure}
  \epsscale{1.0}
 \plotone{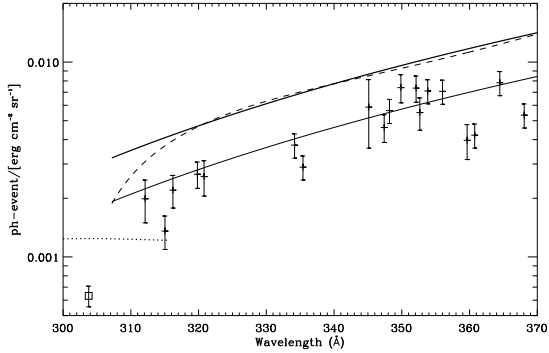}
 \caption{\label{fgcalc}
   The ratio of CDS intensities in detector photon-event units to those measured by EUNIS-06 in absolute units.
The thin solid line represents the updated CDS calibration by EUNIS, obtained by reducing the present SSW 
calibration curve for CDS NIS-1 (thick solid line) by a factor of 1.68 (the average EUNIS-to-CDS 
line ratio). The dotted line represents the present SSW calibration curve for the CDS NIS-2 second order spectra. 
The dashed line represents the SERTS-97 underflight recalibration curve for the CDS NIS-1 \citep{tho02}.  } 
 \end{figure}

 \begin{figure}
 \epsscale{1.0}
 \plotone{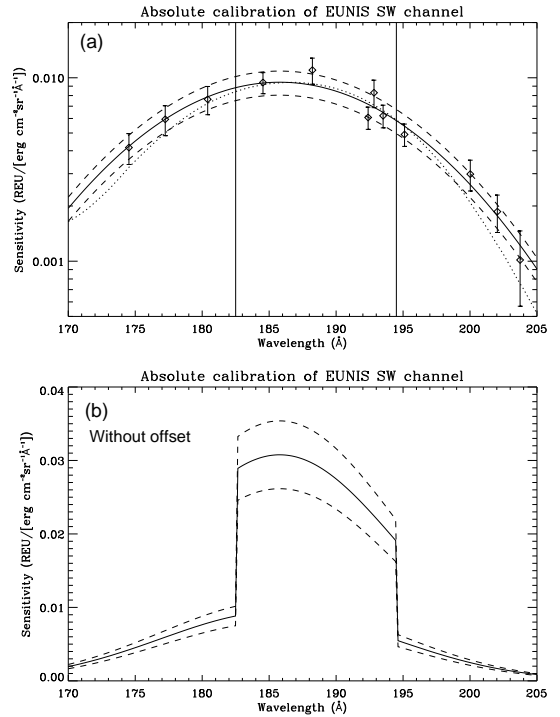}
 \caption{\label{fgswcl}
(a) The measured instrument sensitivity using density- and temperature-insensitive line intensity ratios. The thick
solid curve is a least-square parabolic fit to the data points. The dashed lines indicate the 15\% uncertainty. 
The dotted line represents the relative calibration curve derived by combining measurements of individual 
optical components, which has been scaled to best match the data points. (b) The same calibration curves as 
in (a) but on a linear scale and without correction of relative sensitivity factors for the detectors 
($f_{\rm sensitivity}$ in Eq.~(\ref{eqsen})).}
 \end{figure}

Figure~\ref{fgcalc} shows comparisons between 
the EUNIS-06 underflight recalibration curve and the presently used calibration curves for CDS NIS in SSW. 
Here the EUNIS-06 recalibration curve was obtained by reducing the present SSW calibration curve for CDS NIS-1
by a factor of 1.68 (the average EUNIS-to-CDS line intensity ratio). The units are 
photon-events/(erg cm$^{-2}$sr$^{-1}$), where the term ``photon-events" refers to the photons which have actually 
been detected by the instrument, as opposed to photons which impinge on the detector.

\section{Absolute calibration of EUNIS SW bandpass}
We used the newly calibrated CDS NIS-1 bandpass to calibrate the EUNIS SW channel by means of density-
and temperature-insensitive line intensity ratios because the efficiency of the EUNIS-06 SW bandpass 
could not be directly measured at RAL.
The insensitive line ratio method was proposed by \citet{neu83} as a means of monitoring relative 
calibration variations of inflight EUV spectrometers and was used by \citet{tho94} and \citet{bro96} to adjust 
the laboratory calibration curve for SERTS-89, SERTS-91 and SERTS-93. \citet{bro98a, bro98b} derived the SERTS-95 
relative radiometric calibration for both the first-order and second-order wave bands with this technique. 

The CDS measurements were used because the EUNIS LW and SW channels were not precisely co-pointed in
observation. Figures~\ref{fgspmg}c and \ref{fgspmg}d show the cospatial spectra of EUNIS SW and CDS NIS-1 
bandpasses along their slits. Figures~\ref{fgspfl}a and \ref{fgspfl}b show the spectra averaged over 
just the same spatial area for both instruments. We measured the intensities of
spectral lines for both the EUNIS SW data and the CDS NIS-1 data with a broadened Gaussian fit. The 
measured CDS line intensities were then corrected for calibration update by multiplying a factor 
of 1.68 (the average EUNIS-to-CDS line intensity ratio).

 \begin{figure}
 \epsscale{1.0}  
 \plotone{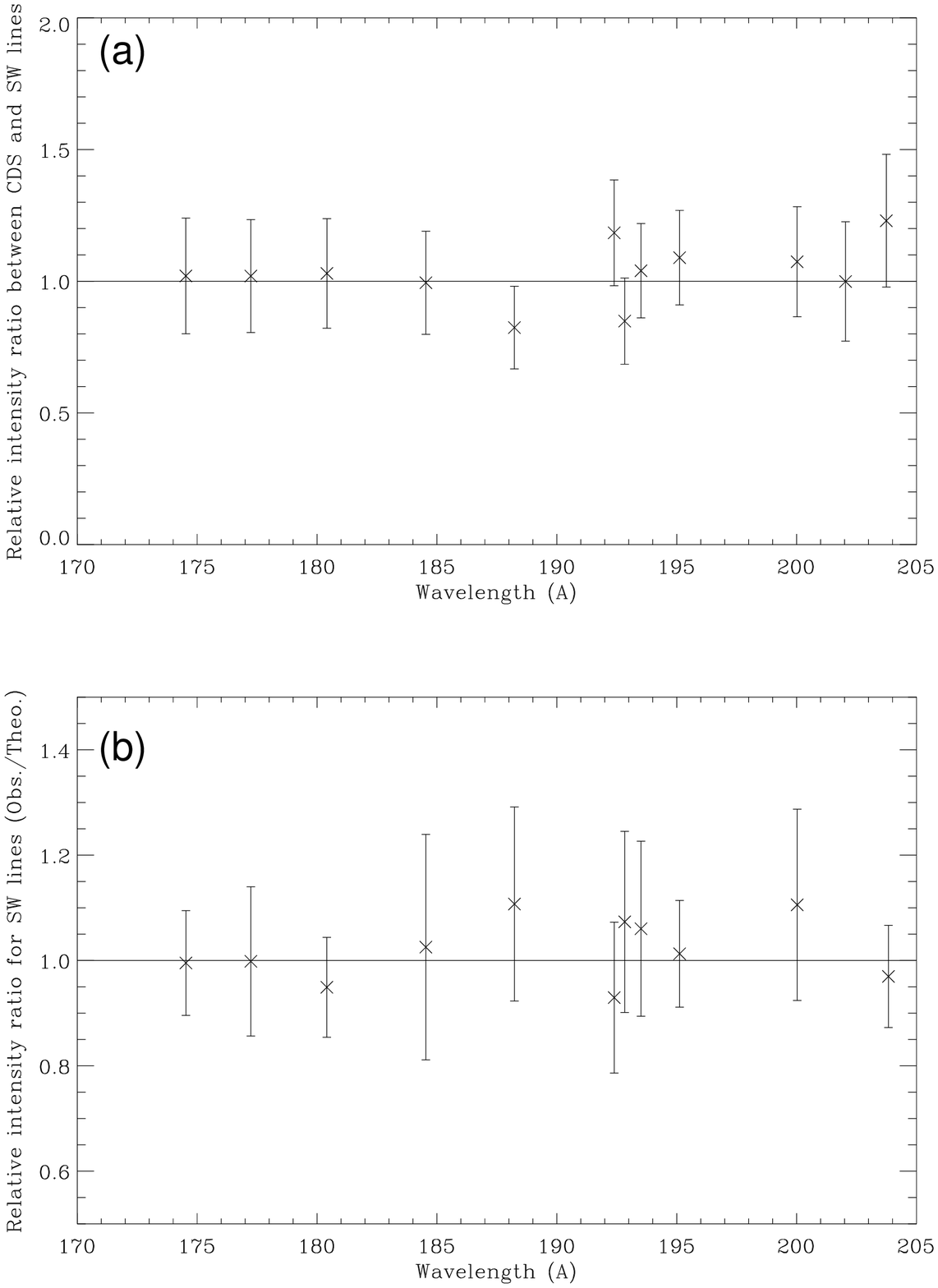}
 \caption{\label{fgtbsw}
 (a) Plot of the observed-to-theoretical intensity ratios between CDS and SW lines (col.[5] of Table~\ref{tabrt})
 as a function of wavelength. (b) Plot of the observed-to-theoretical intensity ratios (col.[6] of 
Table~\ref{tabswr}), normalized to the weighted average ratio within each line group, as a function of wavelength 
for the EUNIS SW channel.}
 \end{figure}

 \begin{deluxetable*}{rllllr}
 \tablecaption{ Density- and temperature-insensitive line groups selected
for EUNIS SW calibration.  The relative line intensity used in calculation of theoretical line ratios 
is in units of erg.  \label{tabrt}}
 \tablewidth{0pt}
 \tablehead{
 \colhead{Ion} &  \colhead{Line Pairs} & \colhead{Theoretical\tablenotemark{a}} & \colhead{Measured\tablenotemark{b}} & \colhead{Col.(4)/Col.(3)} & \colhead{Sensitivity\tablenotemark{c}}\\
 \colhead{(1)} & \colhead{(2)} & \colhead{(3)} & \colhead{(4)} & \colhead{(5)} & \colhead{(6)}}   
 \startdata
Fe\,{\sc{x}}......
& 345.72/174.53  & 0.048$\pm$0.007  & 0.049$\pm$0.008   & 1.020$\pm$0.220   &  4.164$\pm$0.797\\
& 345.72/177.24  & 0.087$\pm$0.012  & 0.089$\pm$0.014   & 1.020$\pm$0.214   &  5.939$\pm$1.105\\
& 345.72/184.54  & 0.201$\pm$0.008  & 0.200$\pm$0.039   & 0.994$\pm$0.196   &  9.428$\pm$1.239\\
\\
Fe\,{\sc{xi}}......
& 352.66/180.41  & 0.082$\pm$0.006  & 0.084$\pm$0.016   & 1.030$\pm$0.208   &  7.628$\pm$1.338\\
& 352.66/188.23  & 0.127$\pm$0.004  & 0.105$\pm$0.020   & 0.824$\pm$0.157   & 11.009$\pm$1.789\\
& 352.66/192.83\tablenotemark{d}  & 0.771$\pm$0.033  & 0.654$\pm$0.123   & 0.849$\pm$0.164   &  8.288$\pm$1.417\\
\\
Fe\,{\sc{xii}}.....
& 352.11/192.39  & 0.311$\pm$0.014  & 0.368$\pm$0.060   & 1.184$\pm$0.201   &  6.075$\pm$0.841\\
& 352.11/193.51  & 0.148$\pm$0.008  & 0.154$\pm$0.025   & 1.040$\pm$0.179   &  6.210$\pm$0.876\\
& 352.11/195.12  & 0.094$\pm$0.002  & 0.102$\pm$0.017   & 1.090$\pm$0.180   &  4.905$\pm$0.683\\
\\
Fe\,{\sc{xiii}}....
& 320.81/200.02  & 0.541$\pm$0.013  & 0.581$\pm$0.112   & 1.074$\pm$0.209   &  2.984$\pm$0.570\\
& 320.81/203.74  & 0.112$\pm$0.009  & 0.138$\pm$0.026   & 1.230$\pm$0.252   &  1.015$\pm$0.447\\
& 348.18/202.04  & 0.148$\pm$0.021  & 0.148$\pm$0.026   & 0.999$\pm$0.227   &  1.864$\pm$0.430\\
 \enddata
\tablenotetext{a}{Theoretical line intensity ratios calculated with the CHIANTI package (ver.6).}
\tablenotetext{b}{The measured CDS to EUNIS SW line intensity ratios, where the EUNIS SW is after
calibration shown in Fig.~\ref{fgswcl}b.}
\tablenotetext{c}{The sensitivity is in units of 10$^{-3}$ REU/[erg cm$^{-2}$sr$^{-1}$\AA$^{-1}$] 
for the SW lines.}
\tablenotetext{d}{Fe\,{\sc{xi}} $\lambda$192.83 is blended with a flare line, Ca\,{\sc{xvii}} $\lambda$192.82
(an EIS ``core line" formed at T=5$\times$10$^6$ K), whose contribution is expected to be small in the present case.}
\end{deluxetable*}

We used the CHIANTI package to obtain theoretical values for density- and 
temperature-insensitive line intensity ratios between 5 emission lines observed by CDS NIS-1 and 12 lines 
observed by EUNIS SW. These include lines from Fe\,{\sc{x}} to Fe\,{\sc{xiii}} listed in Table~\ref{tabrt}.
The first two columns of Table~\ref{tabrt} provide ion name and line pairs, and the third column provides the 
theoretical line intensity ratios. By assuming that the observed line intensity ratio is equal to the 
theoretical one for each line pair, we obtained the sensitivity (or efficiency) of EUNIS SW channel at wavelengths 
for those selected lines. Note that the fiber-optic coupling between the MCP and the APS arrays in each
optical channel leads to differences in the overall sensitivity of each APS array
relative to the other two. However, these differences are uniform over any given APS and are 
easily measured during flat-field testing. After correcting for these factors ($g_i$),
the response of the three APS arrays should be a smoothly varying curve as a function of wavelength. 
Figure~\ref{fgswcl}a shows the obtained sensitivities after this correction [Column (6) of Table~\ref{tabrt}] 
and a least-square parabolic fitting on a logarithmic scale. The absolute calibration response curve 
($f_{\rm sensitivity}$) in the range of 170$-$205\AA\ was obtained by,
\begin{eqnarray}
 f_{\rm sensitivity}& = & g_i 10^{R(\lambda)}, \label{eqsen} \\ 
 R(\lambda)& =& a_0+a_1(\lambda-\lambda_0)+a_2(\lambda-\lambda_0)^2,
\end{eqnarray}   
where $R(\lambda)$ is the fitting function with $\lambda_0$=187.5 \AA, $a_0$=$-$2.03$\pm$0.03, 
$a_1$=$-$(9.5$\pm$2.8)$\times$10$^{-3}$, $a_2$=$-$(2.8$\pm$0.3)$\times$10$^{-3}$, and $g_i$ are the relative 
sensitivity factors for the three APS arrays, given by
\begin{equation}
g_i = \left\{ 
\begin{array}{ll}
1.000 & \quad \mbox{(170 $<\lambda<$182.5  \AA)}\\
3.254 & \quad \mbox{(182.5$<\lambda<$194.5 \AA)}\\
0.950 & \quad \mbox{(194.5$<\lambda<$205  \AA)}\\
\end{array}\right..
\end{equation}
The shape of the response curve derived from solar observations matches well with that of a relative 
calibration curve derived by combining measurements of individual optical components (Fig.~\ref{fgswcl}a). 

 \begin{deluxetable*}{rlcccc}
 \tablecaption{  EUNIS-06 SW channel calibration verification from iron line ratios. \label{tabswr}}
 \tablewidth{0pt}
 \tablehead{
 \colhead{Ion} &  \colhead{Wavelength} & \colhead{Theo. Rel.} & \colhead{Observed Intensity} & \colhead{Observed Rel.} & Normalized\\
  & \colhead{(\AA)} & \colhead{Intensity} & \colhead{(ergs cm$^{-2}$s$^{-1}$sr$^{-1}$)} & \colhead{Intensity} & \colhead{Col.(5)/Col.(3)}\\
 \colhead{(1)} & \colhead{(2)} & \colhead{(3)} & \colhead{(4)} & \colhead{(5)} & \colhead{(6)}}   
 \startdata
Fe\,{\sc{x}} .......
&  174.531   & 1.000$\pm$ 0.000   &2556.87$\pm$ 255.69   & 1.000$\pm$ 0.100   & 0.995$\pm$ 0.100  \\
&  177.240   & 0.550$\pm$ 0.007   &1411.20$\pm$ 141.12   & 0.552$\pm$ 0.078   & 0.998$\pm$ 0.142  \\
&  184.537   & 0.238$\pm$ 0.025   & 626.42$\pm$  93.03   & 0.245$\pm$ 0.044   & 1.025$\pm$ 0.214  \\
\\
Fe\,{\sc{xi}} ......
&  180.408   & 1.000$\pm$ 0.000   &3061.00$\pm$ 306.10   & 1.000$\pm$ 0.100   & 0.949$\pm$ 0.095  \\
&  188.232   & 0.692$\pm$ 0.061   &2469.34$\pm$ 246.93   & 0.807$\pm$ 0.114   & 1.107$\pm$ 0.184  \\
&  192.830   & 0.114$\pm$ 0.009   & 394.99$\pm$  39.50   & 0.129$\pm$ 0.018   & 1.073$\pm$ 0.172  \\
\\
Fe\,{\sc{xii}}  .....
&  192.394   & 0.303$\pm$ 0.019   & 847.44$\pm$  84.74   & 0.278$\pm$ 0.039   & 0.930$\pm$ 0.143  \\
&  193.509   & 0.634$\pm$ 0.043   &2027.08$\pm$ 202.71   & 0.664$\pm$ 0.094   & 1.060$\pm$ 0.166  \\
&  195.119   & 1.000$\pm$ 0.000   &3052.26$\pm$ 305.23   & 1.000$\pm$ 0.100   & 1.013$\pm$ 0.101  \\
\\
Fe\,{\sc{xiii}}  ...
&  200.022   & 0.208$\pm$ 0.015   & 281.62$\pm$  30.45   & 0.237$\pm$ 0.035   & 1.106$\pm$ 0.182  \\
&  203.828   & 1.000$\pm$ 0.000   &1188.14$\pm$ 118.81   & 1.000$\pm$ 0.100   & 0.970$\pm$ 0.097  \\
\enddata
\end{deluxetable*}

We absolutely calibrated the average spectrum of the EUNIS-06 SW channel using the response
function derived above. Figure~\ref{fgspfl}c shows the calibrated SW spectrum, where the background emission has 
been removed. To further demonstrate the validity of SW calibration derived above, we measured absolute intensities 
of the 12 iron lines selected for the calibration from the calibrated spectra by refitting these lines. The resultant
observed line intensity ratios for density- and temperature-insensitive line pairs from CDS NIS-1 and EUNIS SW
are listed in column (4) of Table~\ref{tabrt}. A line-by-line comparison demonstrates good agreement between
theoretical and observed values. This is illustrated in Figure~\ref{fgtbsw}a, where we plot the ratios and 
associated uncertainties of the observed-to-theoretical intensity ratios (col. [5] of Table~\ref{tabrt}) as 
a function of wavelength. The standard deviation of these ratios relative to 1 is about 15\%. The fact that 
no slope is evident in this figure indicates that there is no wavelength bias in the derived absolute radiometric
calibration curve. 

In addition, we examined the relative radiometric calibration of the EUNIS SW channel 
using the same method used for the LW channel.  Table~\ref{tabswr} lists four groups of emission lines 
from Fe\,{\sc{x}} to Fe\,{\sc{xiii}}, in which column (3) gives the theoretical line intensity relative to 
the strongest one in each group. Column (4) gives the observed intensities of calibrated spectral lines measured 
from the averaged spectrum (Fig.~\ref{fgspfl}c), and column (5) the corresponding relative intensities. 
Column (6) gives the ratio of the observed relative intensities to the theoretical ones, further normalized 
by the weighted average ratio within each group. These are plotted in Figure~\ref{fgtbsw}b. 
All of the ratios are equal to unity within their 1$\sigma$ measurement uncertainties, and their maximum deviation 
is less than 15\%. 

Therefore, the results from Figures~\ref{fgtbsw}a and \ref{fgtbsw}b suggest that the uncertainty ($\sigma_1$)
in the relative calibration of the EUNIS SW channel is about 15\%. Other errors that affect our result include 
the 10\% uncertainty ($\sigma_2$) in the absolute calibration of the EUNIS-06 LW channel (see Sect.~\ref{sctlwc}) 
and the typical uncertainty ($\sigma_3$) of about 10\% in the theoretically predicted line intensity ratios used 
in this procedure (estimated from those listed in column (3) of Table~\ref{tabrt}). Thus, we estimate that 
the absolute radiometric calibration of the EUNIS-06 SW channel derived here is accurate to $\pm$20\% overall
(using $\sqrt{\sigma_1^2+\sigma_2^2+\sigma_3^2}$).

\section{Summary}
Using coordinated, cospatial spectroscopic observations of an active region, the lab-calibrated EUNIS LW 
channel has been applied to update the CDS NIS calibration. The improved CDS NIS sensitivity curve was then 
used to derive the EUNIS-06 SW absolute radiometric calibration by applying a technique based on density-
and temperature-insensitive line intensity ratios. Many such ratios of iron lines are mutually consistent
and yield an instrumental response curve that matches well to that derived by combining measurements of 
individual optical components. This result supports the accuracy of the atomic physics parameters. 
Since the EUNIS SW channel has a wavelength range covering wavebands of TRACE and SOHO/EIT, the absolutely
calibrated SW spectra can be used to provide underflight calibration updates for these instruments. In addition,  
data from the second successful EUNIS flight (EUNIS-07, launched on 2007 November 6), will be used to provide 
a calibration update for Hinode/EIS using similar techniques.

\acknowledgments
The EUNIS program is supported by the NASA Heliophysics Division through its Low Cost Access to Space
Program in Solar and Heliospheric Physics. The authors thank the entire EUNIS team for the concerted effort that
led to a successful first flight. TJW is grateful to Dr. William T. Thompson for his valuable comments on 
CDS calibration. The work of TJW was supported by NRL grant N00173-06-1-G033. The authors also thank the
anonymous referee for valuable suggestions. Radiometric calibration of the EUNIS-06 instrument was made 
possible by financial contributions and technical support from both the Rutherford-Appleton Laboratory in 
England and the Physikalisch-Technische Bundesanstalt in Germany, for which we are very grateful.
CHIANTI is a collaborative project involving the NRL (USA), the Universities of Florence (Italy) and 
Cambridge (UK), and George Mason University (USA).

 \clearpage

\end{document}